\documentclass[aps,prb,twocolumn,showpacs,amsmath,amssymb,floatfix,superscriptaddress]{revtex4}
\usepackage{epsfig}
\usepackage{bm}
\usepackage{color}

\usepackage{lipsum}

\begin{document}

\title{Finite frequency noise in chiral Luttinger liquid coupled to phonons}

\author{Edvin G. Idrisov}
\affiliation{Physics and Materials Science Research Unit, University of Luxembourg, L-1511 Luxembourg, Luxembourg}
\date{\today}

\begin{abstract}

We study transport between Quantum Hall (QH) edge states at filling factor $\nu=1$ in the presence of electron-acoustic-phonon coupling. Performing a Bogoliubov-Valatin (BV) trasformation the low-energy spectrum of interacting electron-phonon system is presented. The electron-phonon interaction splits the spectrum into charged and neutral "downstream" and neutral "upstream" modes with different velocities. In the regimes of dc and periodic ac biases the tunelling current and non-equilibrium finite frequency non-symmetrized noise are calculated perturbatively in tunneling coupling of quantum point contact (QPC). We show that the presence of electron-phonon interaction strongly modifies noise and current relations compared to free-fermion case.            
\end{abstract}
 
\maketitle

\section{Introduction}
\label{Sec:I} 

Electron-phonon interactions plays an important role in three-dimensional solid state physics, like for instance to explain superconductivity~\cite{Bardeen1, Bardeen2}. However intense theoretical and experimental investigations have indicated that the static and dynamical properties of electrons in one-dimensional (1D) Luttinger liquids are expected to be strongly modified by the presence of electron-phonon interaction as well~\cite{Kopietz, Sambasiva, Thomas, Solofo, Galda1, Galda2, Galda3, Loss, Martin}. Indeed, using functional bosonization, Galda et. al~\cite{Galda1,Galda2,Galda3} studied the impact of electron-phonon coupling on electron transport through a Luttinger liquid with an embedded scatterer. They investigated the directions of RG flows which can be changed by varying the ratio of Fermi (electron-electron interaction strength) to sound (electron phonon strength) velocities. It was revealed a phase diagram with up to three fixed points: an unstable one with a finite value of conductance and two stable ones, corresponding to an ideal metal or insulator. Next, Martin et. al.~\cite{Loss, Martin} using a Luttinger liquid description calculated the exponents of correlation functions and discussed their remarkable sensitivity to the Wentzel-Bardeen~\cite{Wentzel,Bardeen} singularity induced by the presence of acoustic phonons. Later in Ref.~\cite{Izumida} the transport in suspended metallic single wall carbon nanotube in the presence of strong electron-phonon interaction was investigated. It was shown that the differential conductance as a function of applied bias voltage demonstrates three distinct types of phonon-assisted peaks. These peaks are best observed when the system is in the vicinity of the Wentzel-Bardeen singularity. 

Concurrently, the transport properties of electrons in chiral 1D systems has been a subject of intensive theoretical and experimental studies for a long time as well~\cite{Ezawa}. The experimental investigation of such systems was made possible mostly as a result of discovery of integer and fractional QH effects in two dimensional electron gas (2DEG)~\cite{Klitzing, Tsui}. The chiral electron states appearing at the edge of a 2DEG may be considered as a quantum analogue of classical skipping orbits. Electrons in these 1D chiral states are basically similar to photons: they propagate only in one direction and the is no intrinsic backscattering along the QH edge in case of identical chiralities. However electrons satisfy Fermi-Dirac statistics and are strongly interacting particles.

In order to study chiral 1D edge excitations a low-energy effective theory was introduced by Wen~\cite{Wen1}. The key idea behind this approach is to represent the femionic filed in terms of collective bosonic fields. This so called bozonization technique allows one to diagonalize the edge Hamiltonian of interacting 1D electrons with linear spectrum and calculate any equilibrium correlation functions. Wen's bosonization technique triggered better theoretical understanding of transport properties of 1D chiral systems and explains a lot of experimental findings~\cite{Benoit}. Particularly, the tunneling current, nonequilibrium symmetric quantum noise between chiral fractional QH edge states and photoassisted current and shot noise in fractional QH effect were studied~\cite{Wen, Kivelson, Chamon, Freed, Adeline}. It was shown that the differential conductance is nonzero at finite temperature and manifests a power-law behavior, $G \propto T^{2(g-1)}$, where $T$ is the temperature and $g$ depends on the filling factor of fractional QH state. The exponent $g=\nu$ or $g=1/\nu$ depending on the geometry of tunnelling constriction, where $\nu$ is the filling factor. At zero temperature it has been demonstrated that tunnelling current has the form $I \propto \mu^{2g-1}$, where $\mu$ is an applied dc bias. In lowest order of tunnelling coupling at zero temperature it was shown that the symmetric nonequilibrium noise is given by $S(\omega) \propto \sum_{\sigma=\mp}|\omega+\sigma \omega_0|^{2g-1}$, where $\omega_0=e^{\ast} \mu/\hbar$ is the Josephson frequency of the electron or quasiparticle and $e^{\ast}$ is the effective charge of tunnelling quasi-particle. Thus, an algebraic singularity is present at the Josephson frequency, which depends on charge $e^{\ast}$. At finite temperature it was shown that the ratio between nonequilibrium and equlibrium Johnson-Nyquist noises does not depend on parameter $g$ and has the familiar form $S(\omega,\mu)/S(\omega,0)=(e^{\ast}\mu/2T)\coth(e^{\ast}\mu/2T)$.
 
In Ref.~\cite{Eggert} it has been already shown that chiral QH edge state at filling factor $\nu=1$ is not a Fermi liquid in the presence of electron-phonon interaction. As a consequence the electron correlation function is modified. Particularly, it was demonstrated that ac conductance of chiral channel exhibits resonances at longitudinal wave vectors $q$ and frequencies $\omega$ related by $q=\omega/v_1$,$q=\omega/v_2$ and $q=-\omega/v_3$, where $v_1$,$v_2$, $v_3$ are renormalized Fermi and sound velocities. However it was claimed that the dc Hall conductance is not modified by electron-phonon interaction and is given by $G_H=e^2/2\pi \hbar$. Moreover we would like to mention that according to Wiedemann-Franz law~\cite{Kane, Fisher} the thermal  Hall conductance is not changed as well, namely $K_H=TG_H L_0$, where $T$ is a temperature and $L_0=(\pi^3/3)(k_B/e)^2$ is a free-fermion Lorenz number. 

Motivated by the previous progress in Ref.~\cite{Eggert} we take a further step to investigate the noise properties of tunnelling current between two $\nu=1$ edge states in the presence of strong electron-acoustic phonon interaction. Such a system may be realized in conventional electron optics experiments~\cite{Grenier} and as well in bilayer systems with a total filling factor equal to two~\cite{Ezawa}. At filling factor $\nu=1$ we deal with 1D chiral free-fermion system and no electron-electron interaction influences on tunnelling current and noise. Therefore, any modifications will arise due to electron-phonon interactions. We expect that one may investigate this experimentally as in the works by Milliken et al.~\cite{Milliken}, where the authors measured temperature dependence of tunnelling current between edge states with electron-phonon interaction.

The rest of the paper is organized as follows. In Sec.\ \ref{Sec:II},  we introduce the model of the system, starting with the Hamiltonian of all the constituting parts and the bosonization prescription, and provide a BV transformation in order  to calculate exactly the correlation function.  In Sec.\ \ref{Sec:III}, we calculate perturbatively the tunneling current in the regime of dc bias. In Sec.\ \ref{Sec:IV}, we calculate the finite frequency non-symmetrized noise in the dc regime. Sec.\ \ref{Sec:V} is devoted to the derivation of the tunneling current in the regime of periodic ac bias. The result for finite frequency non-symmetrized noise under the external ac bias are given in Sec.\ \ref{Sec:VI}. We present our conclusions and future perspectives in Sec.\ \ref{Sec:VII}. Details of calculations and the additional information are presented in appendices. Throughout the paper, we set $|e|=\hbar=k_B=1$.

\section{Theoretical model}
\label{Sec:II}
\subsection{QH edge coupled to phonons}
We start by introducing the total Hamiltonian of the system of QH edge states at filling factor $\nu=1$ coupled to acoustic phonons. The relevant energy scales in the experiments with such systems are sufficiently small compared to the Fermi energy, $\epsilon_F$, which suggests using the effective low-energy theory of QH edge states~\cite{Wen}. The advantage of this approach is that it allows to take into account exactly the electron-electron and particularly in our case, the strong electron-phonon interaction. According to the effective theory, edge states can be described as collective fluctuations of the charge density $\hat {\rho}(x)$. The charge density operator is expressed in terms of bosonic field $\hat{\phi}(x)$, namely, $\hat{\rho}(x)=(1/2\pi)\partial_x \hat{\phi}(x)$. The boson field $\hat{\phi}(x)$ can be written in terms of boson creation and annihilation operators, $\hat{a}^{\dagger}_k=\sqrt{L/2 \pi k} \hat{\rho}_k$ and $\hat{a}_k=\sqrt{L/2 \pi k} \hat{\rho}_{-k},$ which satisfy a standard commutation relation $[\hat{a}_k,\hat{a}^{\dagger}_{k^{\prime}}]=\delta_{kk^{\prime}}$, i.e
\begin{equation}
\label{Expansion of bosonic field into creation and annihilation operators}
\hat{\phi}(x)=\hat{\varphi}_0 + 2\pi \hat{\pi}_0 x +\sum_{k>0}\sqrt{\frac{2\pi}{Lk}}[e^{ikx}\hat{a}_k+e^{-ikx}\hat{a}^{\dagger}_k],
\end{equation}
where zero modes fulfils canonical commutation relation $[\hat{\pi}_0,\hat{\varphi}_0]=i/L$ and $L$ is the total size of the system. We consider the thermodynamic limit $L \to +\infty$, consequently $L$ drops out in the final results.

The total Hamiltonian includes three terms  
\begin{equation}
\label{Total Hamiltonian}
\hat{H}=\hat{H}_0+\hat{H}_{ph}+\hat{H}_{e-ph}.
\end{equation} 
Here the first term 
\begin{equation}
\label{Kinetic term of Hamiltonian}
\hat{H}_0=\frac{v_F}{4\pi}\int dx [\partial_x \hat{\phi}(x)]^2
\end{equation}
is the free part of the total Hamiltonian, where $v_F$ is the Fermi velocity.

The second term describes free phonons
\begin{equation}
\label{Phonon term of Hamiltonian}
\hat{H}_{ph}=\frac{1}{2} \int dx \left[\zeta v^2_s (\partial_x \hat{d})^2+\frac{1}{\zeta} \hat{\Pi}^2_d\right],
\end{equation}
where $ \hat{d}(x)$ is a phonon field operator, $\hat{\Pi}_d(x)$ is the canonical conjugate to it, $v_s$ is the sound velocity and $\zeta$ is the linear mass density of the crystal. Here according to Ref.~\cite{Eggert} we disregard the normal modes of phonons perpendicular to the QH edges and consider only the normal mode which is along the QH edge. The phonon field and it's conjugate are given by the following sums of phonon creation and annihilation operators 
\begin{equation}
\label{Phonon and canonical conjugate fields}
\begin{split}
& \hat{\Pi}_d(x)=i\sum_k \sqrt{\frac{v_s \zeta |k|}{2L}} e^{ikx}( \hat{b}^{\dagger}_k-\hat{b}_{-k}), \\
& \hat{d}(x)=\sum_k \sqrt{\frac{1}{2L \zeta v_s|k|}}e^{ikx}(\hat{b}^{\dagger}_k+\hat{b}_{-k}).
\end{split}
\end{equation}
The third term takes into account the strong electron-phonon interaction 
\begin{equation}
\label{Electron-phonon term of Hamiltonian}
\hat{H}_{e-ph}=D \int dx \hat{\rho}(x) \partial_x \hat{d}(x),
\end{equation}
where $D=\sqrt{\pi \zeta v^{3}_s}$ is an electron-phonon coupling constant.  

For further consideration it is convenient to  rewrite the total Hamiltonian in Eq.~(\ref{Total Hamiltonian}) in momentum representation 
\begin{equation}
\label{Total bosonic Hamiltonian in momentum representation}
\begin{split}
& \hat{H}=\sum_{k>0} v_F k \hat{a}^{\dagger}_k \hat{a}_k +\sum_{k>0} v_s k[\hat{b}^{\dagger}_k \hat{b}_k+\hat{b}^{\dagger}_{-k} \hat{b}_{-k}] \\
& +\sum_{k>0} \left[v_c k(\hat{b}^{\dagger}_k+\hat{b}_{-k})\hat{a}_k+h.c. \right],
\end{split}
\end{equation}
where $v_c=D/\sqrt{\pi \zeta v_s}$ and we ignore zero modes as well. For magnetic fields of order of $5$T and 2D electron densities of order of $10^{11}$ cm$^{-2}$, the typical values of velocities for QH edges in GaAs heterojunction are $v_F \simeq v_s$, $v_c/v_s \lesssim 0.1$ and $v_F \simeq 10^5 - 10^6$ cm$/$s~\cite{Ezawa, Glazman}.  

The Hamiltonian in Eq.~(\ref{Total Hamiltonian}) gives the complete description of our system. We note, that Hamiltonian $\hat{H}$ is bilinear in boson creation and annihilation operators, thus the dynamics associated with this Hamiltonian can be accounted exactly by solving linear equations of motion. However, the solution of the  equation of motion has an inconvenient and complex form. To overcome those difficulties, we use the alternative way below applying the BV transformation to diagonalize the total Hamiltonian.
\begin{figure}
\includegraphics[width=8cm]{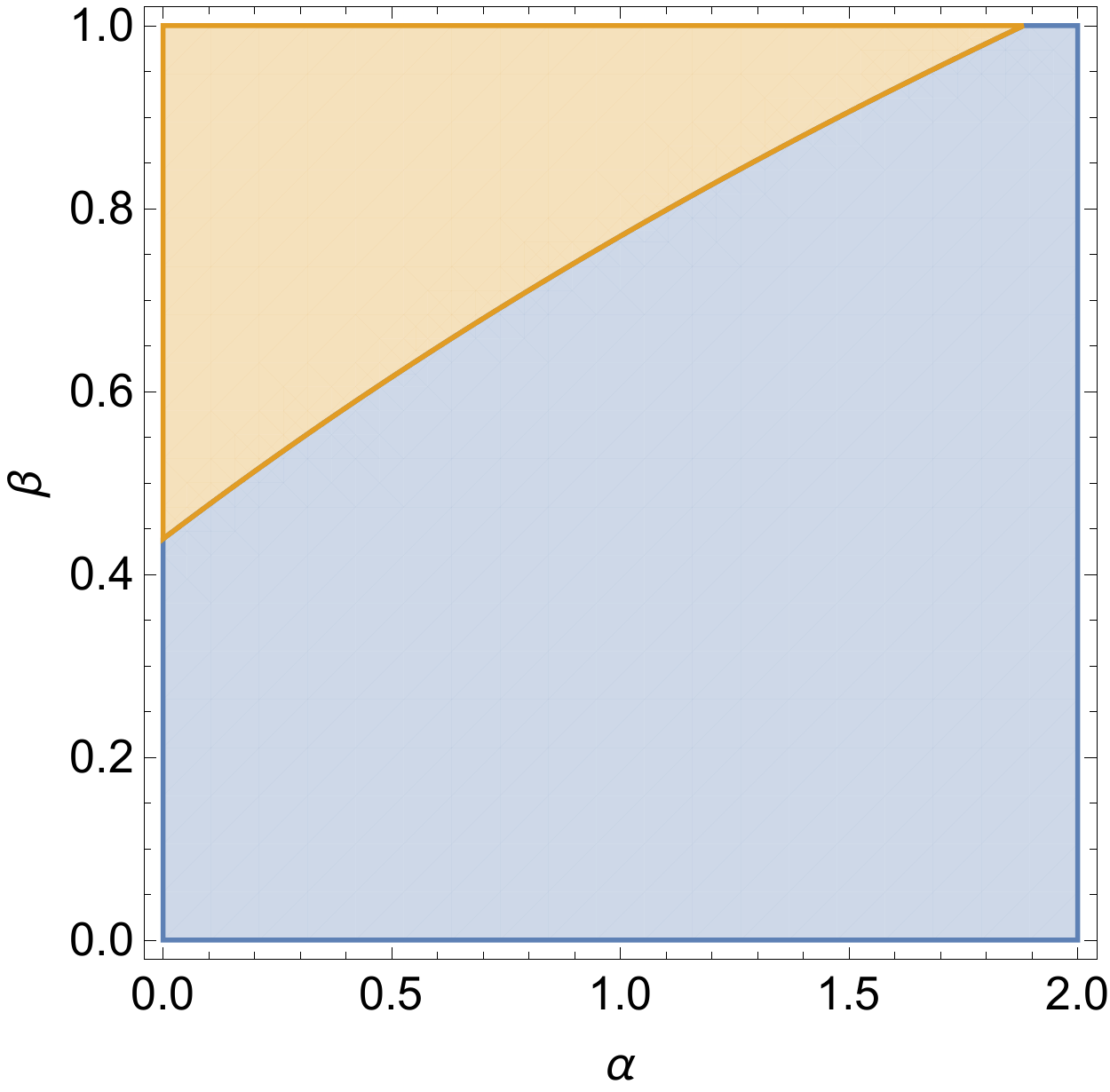}
\caption{\label{fig1} The condition of validation of BV transformation is presented. The light orange region satisfies the inequality $g(\alpha, \beta)<0$ and the light blue region is given by $g(\alpha,\beta)>0$. The BV transformation is valid in light blue region.}
\end{figure}
\begin{figure}
\includegraphics[width=8cm]{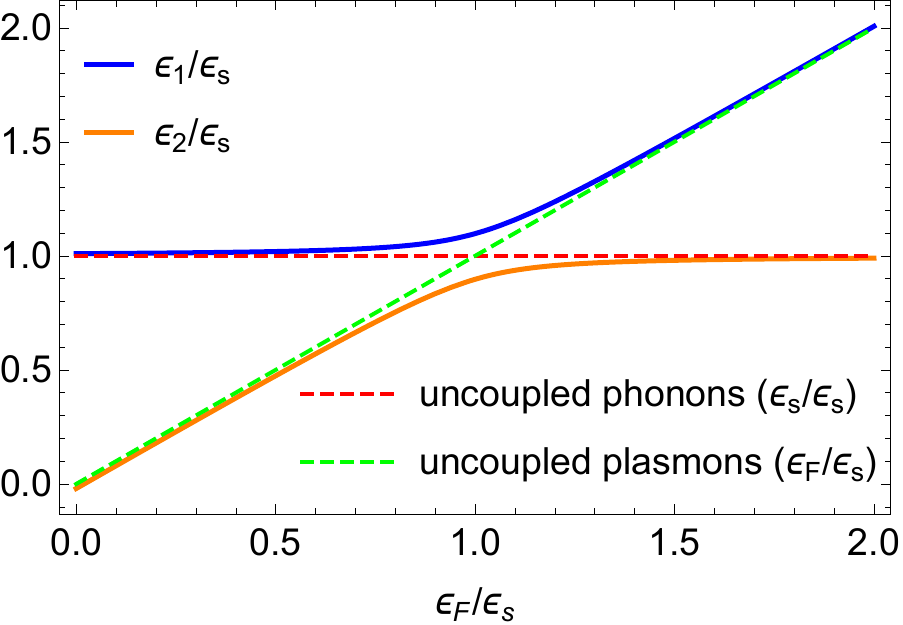}
\caption{\label{fig2.} The renormalized spectrum of $\epsilon_1/\epsilon_s$ and $\epsilon_2/\epsilon_s$ as a function of renormalized Fermi energy $\epsilon_F/\epsilon_s$. For comparison it is shown the spectrum of uncoupled phonon and plazmon system as well. We set $\beta =0.1$.}
\end{figure}
\begin{figure}
\includegraphics[width=8cm]{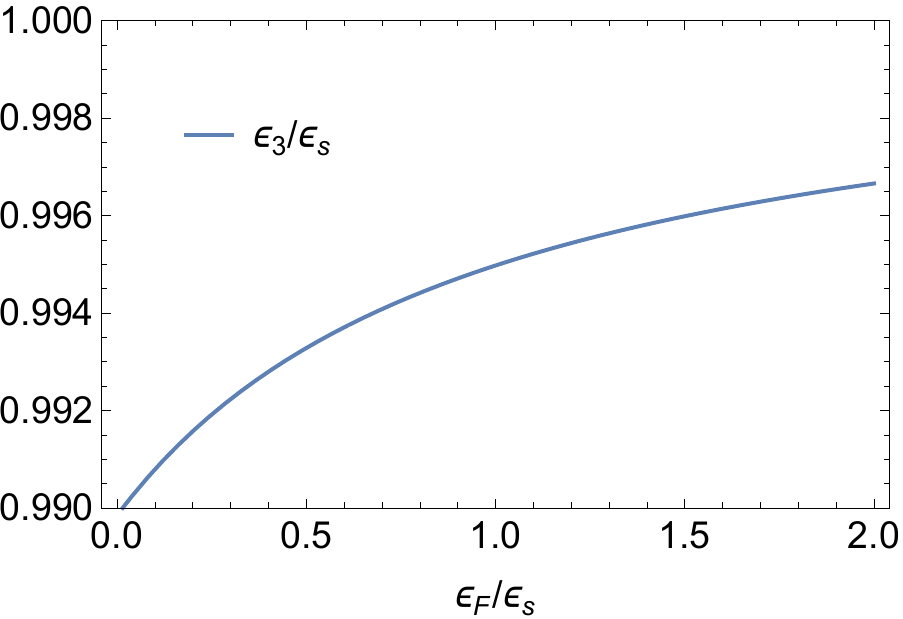}
\caption{\label{fig3.} The renormalized spectrum of $\epsilon_3/\epsilon_s$ as a function of renormalized Fermi energy $\epsilon_F/\epsilon_s$. We set $\beta=0.1$.}
\end{figure}  
\begin{figure}
\includegraphics[width=8cm]{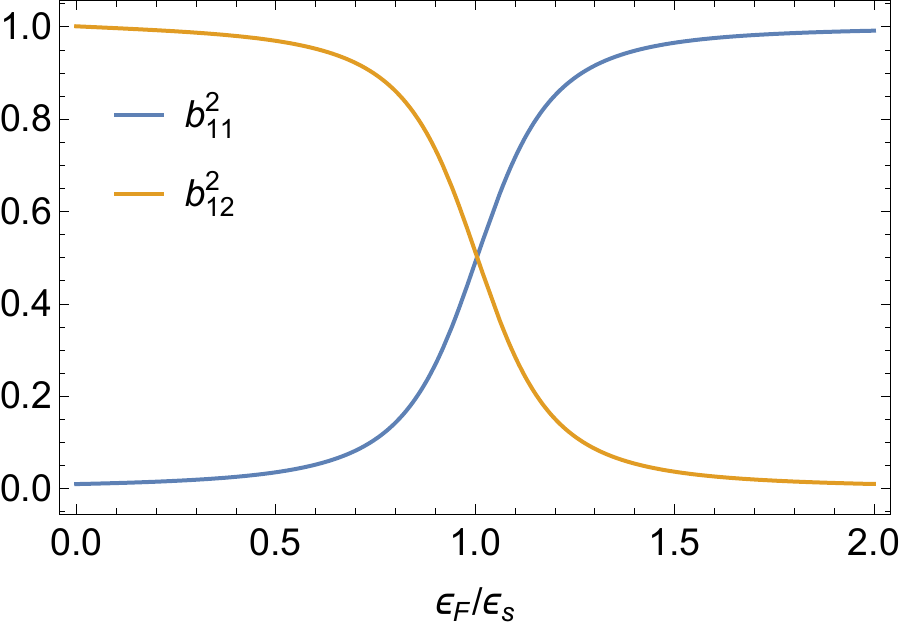}
\caption{\label{fig4.} The squares of first two matrix elements of BV transformation as a function of renormalized Fermi energy $\epsilon_F/\epsilon_s$ (see Eq.~(\ref{BV matrix})). We set $\beta=0.1$.}
\end{figure}

\subsection{Bogoliubov-Valatin transformation}
Using operator identity $[\hat{a} \hat{b}, \hat{c}]=\hat{a}[\hat{b},\hat{c}]+[\hat{a},\hat{c}]\hat{b}$ for bosonic operators, the equations of motion can be derived from Eq.~(\ref{Total bosonic Hamiltonian in momentum representation})
\begin{equation}
\label{Equation of motion}
\frac{d}{dt}
\begin{bmatrix}
  \hat{a}_k(t) \\
  \hat{b}_k(t)  \\
  \hat{b}^{\dagger}_{-k}(t)
 \end{bmatrix}
 =-iv_s k A
\begin{bmatrix}
  \hat{a}_k(t) \\
  \hat{b}_k(t)  \\
  \hat{b}^{\dagger}_{-k}(t)
 \end{bmatrix}.
\end{equation}
Here the dimensionless $3 \times 3$ matrix has the following form
\begin{equation}
\label{Dynamical matrix}
A=
\begin{bmatrix}
       \alpha & \beta & \beta \\
       \beta & 1 & 0 \\
       -\beta & 0 & -1
     \end{bmatrix}
\end{equation}
where $\alpha=v_F/v_s$ and $\beta=v_c/v_s$. We call it a dynamical matrix.

Next according to Ref.~\cite{Ming}, the Hamiltonian of bosons in Eq.~(\ref{Total bosonic Hamiltonian in momentum representation}) is BV diagonalizable if it's dynamical matrix is physically diagonalizable. The dynamical matrix said to be physically diagonalizable if it is diagonalizable, and all its eigenvalues are real. Therefore we write the characteristic equation to find eigenvalues
\begin{equation}
\label{Characteristic equation}
\lambda^3-\alpha \lambda^2-\lambda+\chi=0,
\end{equation}
where we introduced the notation $\chi=\alpha-2\beta^2$. To have three distinct real eigenvalues the coefficients of characteristic cubic equation must satisfy the inequality (see Fig.~\ref{fig1})   
\begin{equation}
\label{Inequality}
g(\alpha, \beta)= 18 \alpha \chi+4\alpha^3 \chi+(\alpha^2+4)-27\chi^2>0.
\end{equation}
The eigenvalues of dynamical matrix~(\ref{Dynamical matrix}) are provided in Appendix~\ref{Sec:A}. Using the ideas of Ref.~\cite{Ming} one can construct the BV transformation matrix 
\begin{widetext}
\begin{equation}
\label{BV matrix}
B=
\begin{bmatrix}
  b_{11} & b_{12} & b_{13} \\
  b_{21} & b_{22} & b_{23} \\
  b_{31}  & b_{32}  & b_{33}  
 \end{bmatrix}
= 
\begin{bmatrix}
  \sqrt{\frac{(\lambda^2_1-1)^2}{(\lambda^2_1-1)^2+4\beta^2 \lambda_1}} &  \sqrt{\frac{(\lambda^2_2-1)^2}{(\lambda^2_2-1)^2+4\beta^2 \lambda_2}} &  \sqrt{\frac{(\lambda^2_3-1)^2}{-(\lambda^2_3-1)^2-4\beta^2 \lambda_3}} \\[0.3em]
 \frac{\beta}{\lambda_1-1} \sqrt{\frac{(\lambda^2_1-1)^2}{(\lambda^2_1-1)^2+4\beta^2 \lambda_1}}&  \frac{\beta}{\lambda_2-1}\sqrt{\frac{(\lambda^2_2-1)^2}{(\lambda^2_2-1)^2+4\beta^2 \lambda_2}} &  \frac{\beta}{\lambda_3-1} \sqrt{\frac{(\lambda^2_3-1)^2}{-(\lambda^2_3-1)^2-4\beta^2 \lambda_3}}\\[0.3em]
\frac{-\beta}{\lambda_1+1} \sqrt{\frac{(\lambda^2_1-1)^2}{(\lambda^2_1-1)^2+4\beta^2 \lambda_1}}&  \frac{-\beta}{\lambda_2+1}\sqrt{\frac{(\lambda^2_2-1)^2}{(\lambda^2_2-1)^2+4\beta^2 \lambda_2}} &  \frac{-\beta}{\lambda_3+1} \sqrt{\frac{(\lambda^2_3-1)^2}{-(\lambda^2_3-1)^2-4\beta^2 \lambda_3}}\\
 \end{bmatrix},
\end{equation}
\end{widetext}
where $\lambda_1$, $\lambda_2$ and $\lambda_3$ are real eigenvalues of Eq.~(\ref{Characteristic equation}).
\begin{figure}
\includegraphics[width=8cm]{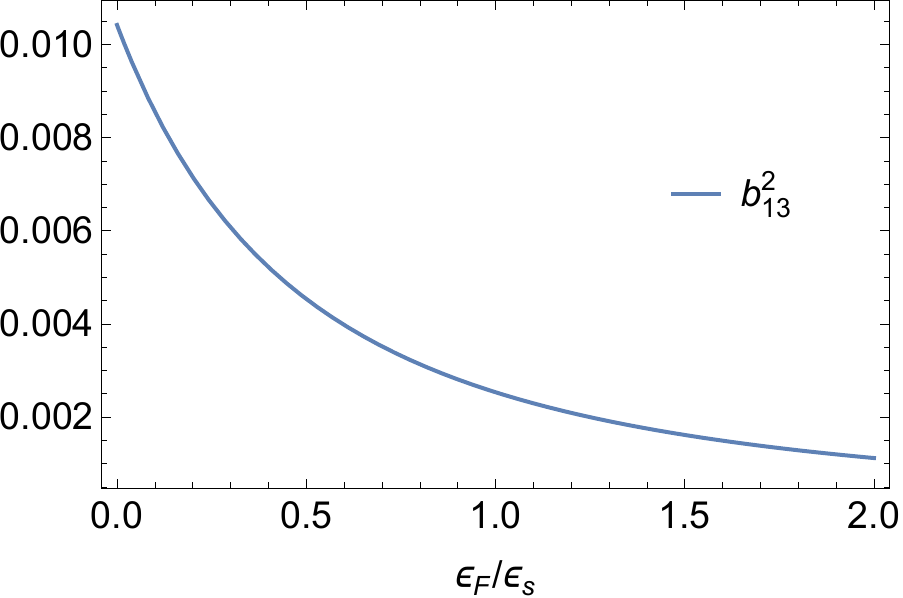}
\caption{\label{fig5.} The square of third matrix element of BV transformation as a function of renormalized Fermi energy $\epsilon_F/\epsilon_s$ (see Eq.~(\ref{BV matrix})). We set $\beta=0.1$.}
\end{figure}
\begin{figure}
\includegraphics[width=8cm]{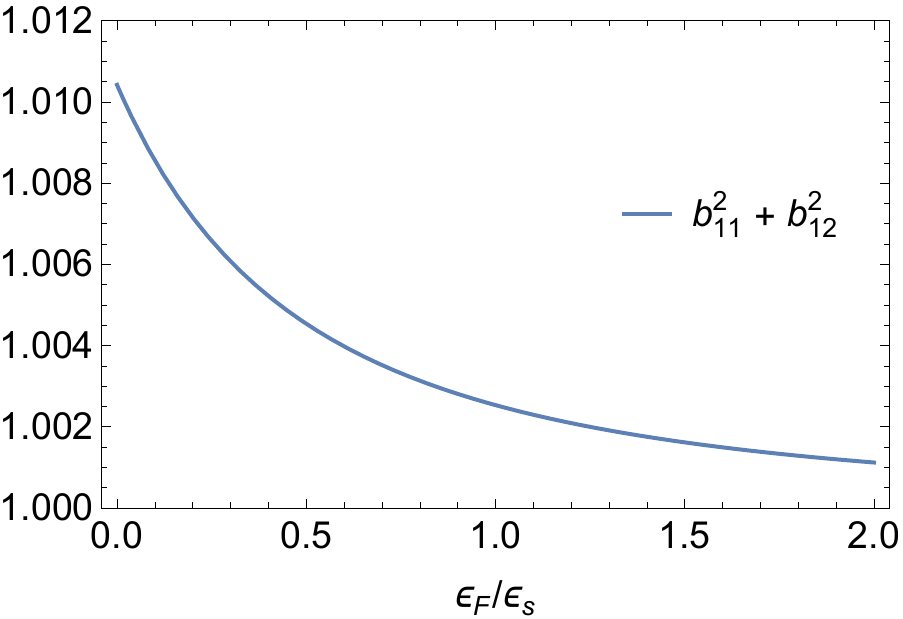}
\caption{\label{fig6.} The sum of squares of first two BV matrix elements.(See Eq.~(\ref{BV matrix})). We set $\beta=0.1$. Note, that the sum is not equal to one.}
\end{figure}

\begin{figure}
\includegraphics[width=8cm]{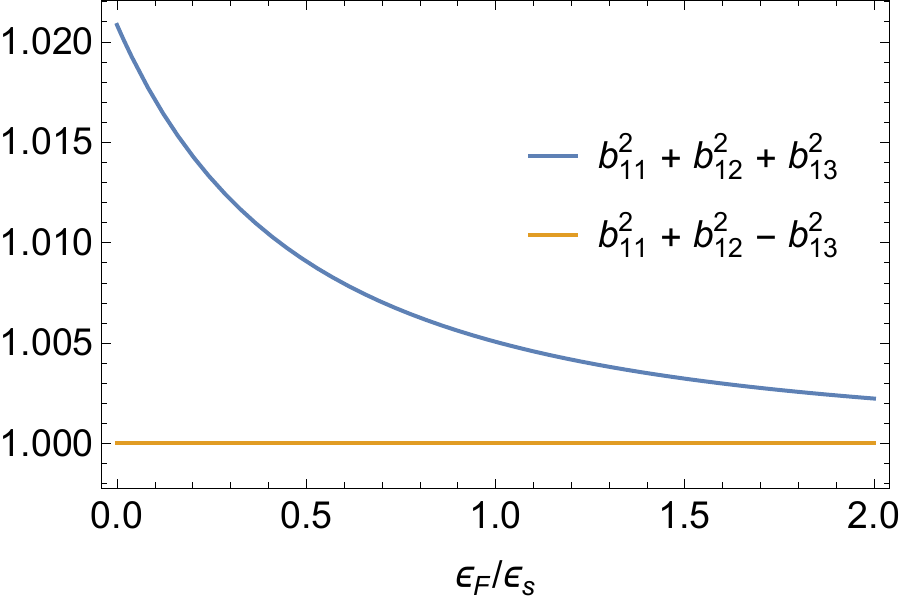}
\caption{\label{fig7.} It is shown the relations between first row elements of BV matrix in Eq.~(\ref{BV matrix}). We set $\beta=0.1$.}
\end{figure}

The straightforward algebra confirms that the BV matrix diagonalizes the initial total bosonic Hamiltonian~(\ref{Total bosonic Hamiltonian in momentum representation}). Namely, we get the following quadratic form (the nonessential constant term is omitted) 
\begin{equation}
\label{New diagonalized Hamiltonian}
\hat{H}=\sum_{k>0} \epsilon_1 \hat{\alpha}^{\dagger}_k \hat{\alpha}_k + \sum_{k>0}\epsilon_2 \hat{\beta}^{\dagger}_k \hat{\beta}_k +\sum_{k>0}\epsilon_3 \hat{\gamma}^{\dagger}_{-k}\hat{\gamma}_{-k},
\end{equation}
where $\epsilon_i = v_i k$, $i=1,2,3$ and $v_1=\lambda_1 v_s>0$, $v_2=\lambda_2 v_s>0$, $v_3=-\lambda_3 v_s>0$. One can check that the relations $\epsilon_1+\epsilon_2-\epsilon_3=\epsilon_F$ and $b^2_{11}+b^2_{12}-b^2_{13}=1$ are invariants of BV transformation and the statistics of system, i.e commutation relations of boson operators, is preserved after the BV transformation. The renormalized spectrum is given in Fig.~\ref{fig2.} and Fig.~\ref{fig3.}. From these figures we see that $\epsilon_1$, $\epsilon_2$ are strongly modified by electron-phonon interactions in comparison to free plasmons$-$ collective excitation of QH edge and free phonons, and $\epsilon_3$ is slightly changed from it's initial value $\epsilon_s$. The elements of first row of BV transformation are given in Figs.~\ref{fig4.}$-$\ref{fig7.}.

According to the BV transformation the bosonic operator is expressed in terms of three independent bosonic terms, namely 
\begin{equation}
\label{Bosonic operator after BV transformation}
\hat{a}_k=b_{11}\hat{\alpha}_k+b_{12} \hat{\beta}_k+b_{13} \hat{\gamma}^{\dagger}_{-k}
\end{equation}
This equation is the main result of this subsection. It will be used in next subsection to calculate the two-point correlation function. 

\subsection{Correlation function}
The important quantity which allows to describe the transport properties of system under consideration is equilibrium two-point correlation function. It is defined as $G(x_1,t_1;x_2,t_2)=\langle \hat{\psi}^{\dagger}(x_1,t_1) \hat{\psi}(x_2,t_2)\rangle$, where according to bosonization technique the fermionic field is given by $ \hat{\psi}(x)\propto \exp[i \hat{\phi}(x)]$, where we omit the ultraviolet cut-off prefactor. The average is taken with respect to equilibrium density matrix $ \hat{\rho}_0 \propto \exp[( \hat{H}_0+\hat{H}_{ph})/T]$ and $T$ is temperature.

Next using the expansion of bosonic fields in terms of collective modes in Eq.~(\ref{Expansion of bosonic field into creation and annihilation operators}) and the BV transformation in Eqs.~(\ref{BV matrix}-\ref{Bosonic operator after BV transformation}) we finally arrive to the following result for the fermion correlation function (see Appendix~\ref{Sec:B})

\begin{widetext}
\begin{equation}
\label{Correlation function at zero temperature}
G(x_1,t_1;x_2,t_2) \propto \left[\frac{i\eta/2\pi}{x_1-x_2-v_1(t_1-t_2)+i\eta}\right]^{b^2_{11}}
\left[\frac{i\eta/2\pi}{x_1-x_2-v_2(t_1-t_2)+i\eta}\right]^{b^2_{12}} \left[\frac{-i\eta/2\pi}{x_1-x_2+v_3(t_1-t_2)-i\eta}\right]^{b^2_{13}}.
\end{equation}
\end{widetext}

The free-fermion two point correlation function for right moving mode is proportional to $\propto (i\eta/2\pi)/(x_1-x_2-v_F(t_1-t_2)+i\eta)$. Here we see that the electron-phonon interaction changes the correlation function sufficiently compared to free fermion one. 

The finite temperature correlation function is obtained applying a conformal transformation~\cite{Tsvelik}. To apply the conformal transformation we go to Euclidean time $-i\tau=t_1-t_2-(x_1-x_2)/v$ and then set $v \tau \to (v/2\pi T)\arctan(2\pi T \tau)$. It follows that 
\begin{widetext}
\begin{equation}
\label{Correlationfunction at finite temperature}
\begin{split}
& G(x_1,t_1;x_2,t_2) \propto \left[\frac{i\eta T/2v_1}{\sinh[\frac{\pi T}{v_1}(x_1-x_2-v_{\alpha}(t_1-t_2))+i\eta]}\right]^{b^2_{11}} \times \\
& \left[\frac{i\eta T/2v_2}{\sinh[\frac{\pi T}{v_2}(x_1-x_2-v_2(t_1-t_2))+i\eta]}\right]^{b^2_{12}} \left[\frac{-i\eta T/2v_3}{\sinh[\frac{\pi T}{v_3}(x_1-x_2+v_3(t_1-t_2))-i\eta]}\right]^{b^2_{13}}.
\end{split}
\end{equation}
\end{widetext}

As one can mention, the correlation function splits into three independent correlation functions with different velocities. Two of them with velocities $v_1$, $v_2$ correspond to "downstream" modes. One of this modes, $v_1$, carries a charge. The third correlation function is related to the "upstream" mode, and travels in the opposite side with respect to right-moving "downstream" modes. In the following sections we use these correlators to derive the tunneling current and noise perturbatively in tunneling coupling.
\section{Tunneling current in dc regime}
\label{Sec:III}
\subsection{Tunneling Hamiltonian}
The electron-phonon interaction is taken into account exactly therefore the tunneling has to be considered perturbatively. The tunneling (backscattering) Hamiltonian of electrons at the QPC located at point $x$, is given by  
\begin{equation}
\label{Tunneling Hamiltonian}
\hat{H}_{T}= \hat{A}+\hat{A}^{\dagger}, \quad \hat{A}= \tau \hat{\psi}^{\dagger}_d(x) \hat{\psi}_u(x),
\end{equation}
where $\tau$ (generally a complex number) is the tunneling coupling constants, $ \hat{\psi}(x)$ is an electron annihilation operator~\cite{Slobodeniuk}. The vertex operator $ \hat{A}$ in Eq.~(\ref{Tunneling Hamiltonian}) mixes the up ($u$) and down ($d$) channels and transfers electrons from one channel to another.  

We introduce the tunneling current operator $\hat{J}=\dot{\hat{N}}_d=i[\hat{H}_T,\hat{N}_d]$, where $\hat{N}_d=\int dx \hat{\psi}^{\dagger}_d(x) \hat{\psi}_d(x)$ is the number of electrons in the down channel. After simple algebra we obtain the tunneling current operator 
\begin{equation}
\hat{J}=i(\hat{A}^{\dagger}-\hat{A}).
\end{equation}
In the interaction representation the average current is given by the expression
\begin{equation}
\label{Average current. Symmetric barriers}
I = \langle \hat{U}^{\dagger}(t,-\infty)\hat{J}(t) \hat{U}(t,-\infty) \rangle,
\end{equation}   
where the average is taken with respect to dc biased ground state in QH edges and free phonons and 
\begin{equation}
\label{Evolutional operator. Symmetric barriers}
\hat{U}(t_1,t_2) = \hat{\text{T}}\text{exp}\left[-i\int_{t_2}^{t_1} dt \hat{H}_{T}(t)\right]
\end{equation} 
is the evolution operator.

We evaluate the average current perturbatively expanding the evolution operator to the lowest order in tunneling amplitude. It is obvious to observe that the average tunneling current can be written as a commutator of vertex operators, namely one obtain
\begin{equation}
\label{Tunneling current at DC bias}
I = \int dt \langle [\hat{A}^{\dagger}(t), \hat{A}(0)]\rangle.
\end{equation}
In our model there is no interaction between upper and lower channels, i.e they are independent and therefore the correlation functions in Eq.~(\ref{Tunneling current at DC bias}) splits into the product of two single-particle correlators, namely 
\begin{widetext}
\begin{equation}
\label{Tunneling current at dc bias 2}
I = |\tau|^2 \int dt e^{i \mu t}\left[\langle \hat{\psi}^{\dagger}_u (t)\hat{\psi}_u(0)\rangle \langle \hat{\psi}_d (t)\hat{\psi}^{\dagger}_d(0)\rangle -  \langle \hat{\psi}^{\dagger}_d (0)\hat{\psi}_d(t)\rangle \langle \hat{\psi}_u (0)\hat{\psi}^{\dagger}_u(t)\rangle \right],
\end{equation}
\end{widetext}
where we set $x=0$, the dc bias $\mu$ is applied to upper chiral channel and the average now is taken with respect to equilibrium density matrix $\hat{\rho}_0$ from Sec.~\ref{Sec:II}. 
\begin{figure}
\includegraphics[width=8cm]{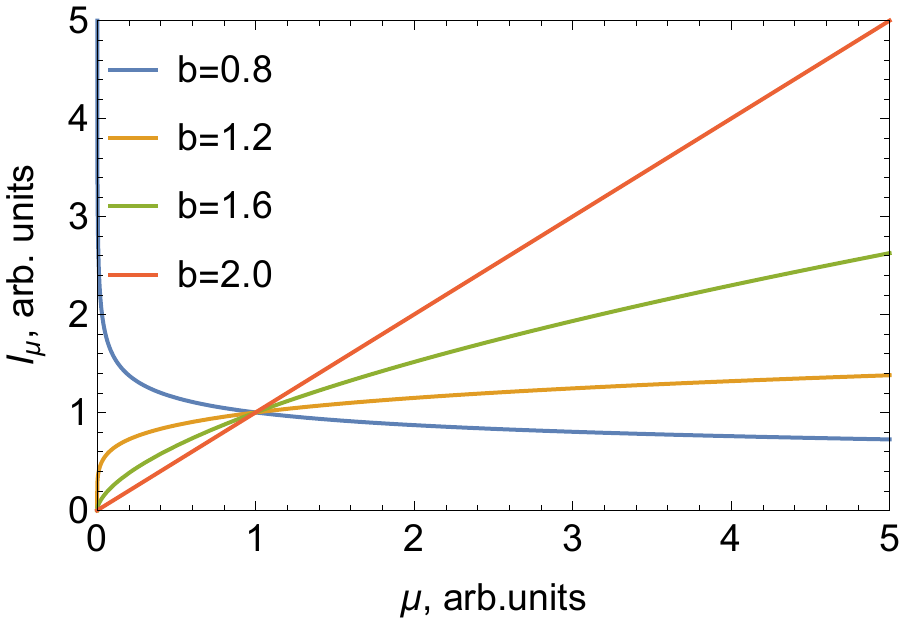}
\caption{\label{fig8.} The tunneling current $I_{\mu}$ versus applied dc voltage $\mu$ at zero temperature for different even values of $b$ is shown.}
\end{figure}
Next without loss of generality we consider the case of positive bias $\mu>0$. The sign of bias $\mu$ simply determines the direction of current. Substituting the correlation functions from Eq.~(\ref{Correlation function at zero temperature}) into Eq.~(\ref{Tunneling current at dc bias 2}) and employing the integral
\begin{equation}
\lim_{\eta \to +0}\int \frac{e^{\eta+ix}}{(\eta+ix)^b} dx= \frac{2\pi}{\Gamma[b]}, 
\end{equation}
where $x=\mu t$ is a dimensionless variable of integration, we obtain the expression for tunneling current at zero temperature 
\begin{equation}
\label{Tunneling current at zero temperature. Result.}
I_{\mu} \propto \mathcal{D} |\tau|^2 \frac{2\pi}{\Gamma[b]} \mu^{b-1},
\end{equation}
where $\mathcal{D}=(2\pi v_1)^{-2b^2_{11}}(2\pi v_2)^{-2b^2_{12}}(2\pi v_3)^{-2b^2_{13}}$ is the renormalized density of states in presence of electron-phonon interaction and $b=2(b^2_{11}+b^2_{12}+b^2_{13})$, $\Gamma[b]$ is the Euler gamma function and we omit the prefactor which depends on ultraviolet cut-off. The dependence of tunneling current on applied dc bias is presented in Fig.~\ref{fig8.} and Fig.~\ref{fig9.}. It is a monotonic function.

For the case of finite temperature an analytical continuation on complex plane is applied to Eq.~(\ref{Tunneling current at dc bias 2}). Then shifting the integration variable $\pi T t=u+i\pi/2$ (this does not affect the singularities of integrand) and using the integral
\begin{equation}
\lim_{\eta \to +0}\int \frac{e^{i\frac{\mu}{\pi T}u}}{\cosh^{b}[u+i\eta]} du= 2^{b-1}\frac{\left|\Gamma\left[\frac{1}{2}(b+i\frac{\mu}{\pi T})\right]\right|^2}{\Gamma[b]},
\end{equation}
we get the final result for tunneling current
\begin{equation}
\label{Tunneling current at finite temperature. Result.}
I \propto 2 \mathcal{D} |\tau|^2 (2\pi T)^{b}\sinh\left(\frac{\mu}{2T}\right)\frac{\left|\Gamma\left[\frac{1}{2}(b+i\frac{\mu}{\pi T})\right]\right|^2}{2\pi T\Gamma[b]}.
\end{equation}
In the non-interacting free-fermion case at $b=2$, the current is given by the well-known Landauer-Buttiker formula $ I= |\tau|^2 \mu /2\pi v_F$ and QPC is in ohmic regime.
\begin{figure}
\includegraphics[width=8cm]{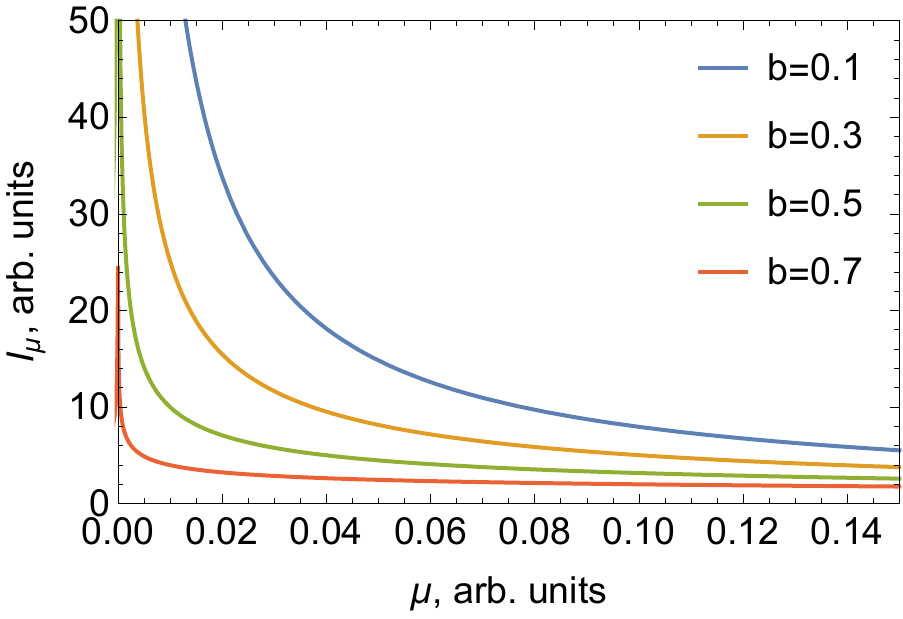}
\caption{\label{fig9.} The tunneling current $I_{\mu}$ versus applied dc voltage $\mu$ at zero temperature for different odd values of $b$ is shown.}
\end{figure}
However a non-ohmic behavior could be observed for non-Fermi liquid. Namely, at high temperatures from Eq.~(\ref{Tunneling current at finite temperature. Result.}) we obtain
\begin{equation}
I \propto T^{b-2} \mu, \quad \mu/T \ll 1,
\end{equation}
which demonstrates an ohmic behavior, and at low temperatures we get the expression
\begin{equation}
 I \propto \mu^{b-1}, \quad \mu/T \gg 1, 
\end{equation}
with non-ohmic behavior, specific for non-Fermi liquid.

Next apart from current or conductance it is useful to measure the time dependent fluctuations of tunneling current, noise. For instance, it is known that measuring shot noise allows a direct access to fractional charges of Laughlin quasiparticles~\cite{Chamon}. The calculation of noise is the subject of next sections.   
 
\section{Finite and zero frequency noise in dc regime}
\label{Sec:IV}
\subsection{Finite frequency noise}
Experimentally the measured spectral noise density of current fluctuations strongly depends on how the detector operates~\cite{Lesovik}. We consider the finite frequency non-symmetrized noise. In the absence of time-dependent external fields the correlation function in Eqs.~(\ref{Correlation function at zero temperature}) and (\ref{Correlationfunction at finite temperature}) depends on the time difference. This is true as well for finite frequency noise defined as 
\begin{equation}
\label{Definition of noise dc bias}
S(\omega)= \int dt e^{i\omega t} \langle \delta \hat{J}(t) \delta \hat{J}(0)\rangle,
\end{equation}
where the average is taken with respect to dc biased ground state in QH edges and free phonons as in Eq.~(\ref{Tunneling current at DC bias}) and $\delta \hat{J}(t)=\hat{J}(t)-\langle \hat{J}(t)\rangle$. This has become a measurable quantity in recent experiments~\cite{Basset}. We evaluate the Eq.~(\ref{Definition of noise dc bias}) perturbatively in the lowest order of tunneling coupling and get the following expression for finite-frequency non-symmetrized noise in terms of vertex operators, defined in Eq.~(\ref{Tunneling Hamiltonian}), namely 
\begin{equation}
\label{Finite frequency noise in terms of vertex operators in dc regime}
S(\omega)=\int dt e^{i\omega t} [\langle \hat{A}^{\dagger}(t)\hat{A}(0)\rangle + \langle \hat{A}(t)\hat{A}^{\dagger}(0)\rangle].
\end{equation}
One can mention that to obtain the symmetrized noise one need simply construct an even function of non-symmetrized noise, i.e $[S(\omega)+S(-\omega)]/2$.  

\begin{figure}
\includegraphics[width=8cm]{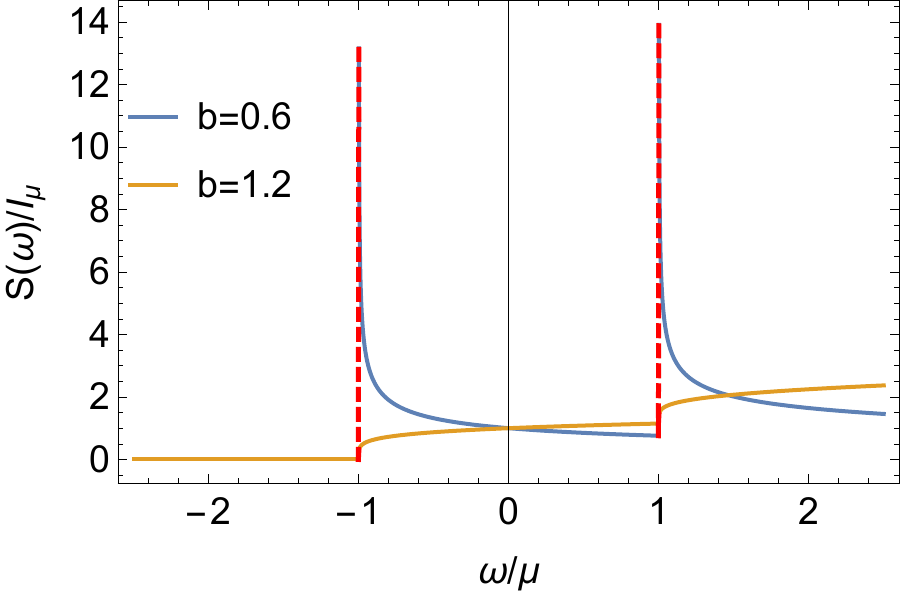}
\caption{\label{fig10.} The normalized non-symmetrized finite frequency noise $S(\omega)/I_{\mu}$ is plotted versus the dimensionless frequency $\omega / \mu$ at zero temperature and fixed dc bias $\mu$. We set $b=0.6$ and $b=1.2$.}
\end{figure}

Evaluating the Eq.~(\ref{Finite frequency noise in terms of vertex operators in dc regime}) at zero temperature we obtain the result for finite frequency non-symmetrized noise
\begin{equation}
\label{Result for finite frequency noise at dc bias and zero temperatire}
S(\omega) \propto \mathcal{D}|\tau|^2 \frac{2\pi}{\Gamma[b]} \sum_{\sigma=\mp}|\omega+\sigma \mu|^{b-1}\theta(\omega+\sigma\mu),
\end{equation} 
where $\theta(x)$ is the Heaviside function. The normalized finite frequency non-symmetrized noise is plotted in Fig.~\ref{fig10.}. From Eq.~(\ref{Result for finite frequency noise at dc bias and zero temperatire}) one can notice that at $\omega \ll \mu$, we obtain $S(\omega) \propto I_{\mu}$, which is the classical shot noise result of next subsection. As expected it does not depend on the interaction parameter $b$. Note that at $\omega = \mp \mu$ there are knife-edged singularities in finite frequency noise. In case of non-interacting fermionic liquid at $b=2$ these singularities are caused by Pauli exclusion principle~\cite{Levitov}.  

For finite temperatures we substitute the correlation function from Eq.~(\ref{Correlationfunction at finite temperature}) into Eq.~(\ref{Finite frequency noise in terms of vertex operators in dc regime}) and get 
\begin{equation}
\label{Result for finite frequency noise at dc bias and finite temperatire}
S(\omega) \propto \mathcal{D}|\tau|^2 (2\pi T)^{b}\sum_{\sigma=\mp}\frac{\exp(\frac{\omega+\sigma \mu}{2T})\left|\Gamma\left[\frac{1}{2}(b+i\frac{\omega+\sigma \mu}{\pi T})\right]\right|^2}{2\pi T\Gamma[b]}.
\end{equation}
Away from singularities at $|\omega \mp \mu|/2T \gg 1$ we recover the Eq.~(\ref{Result for finite frequency noise at dc bias and zero temperatire}). At zero frequency this result coincides with the Eq.~(\ref{Shot noise in dc regime}) of next subsection. 

\subsection{Zero frequency noise}
Experimentally, usually it is the zero frequency spectral noise density which is studied. Moreover, in the limit of zero frequency the non-symmetrized and symmetrized noises coincide. Unlike tunneling current, the zero frequency noise is given by the anti-commutator of vertex operators  
\begin{equation}
S(0)=\int dt \langle \{ \hat{A}^{\dagger}(t), \hat{A}(0)\}\rangle.
\end{equation}
A straightforward calculations similar to what we have done for evaluating the tunneling current in Eq.~(\ref{Tunneling current at finite temperature. Result.}) gives
\begin{equation}
\label{Shot noise in dc regime}
S(0)/I = \coth \left( \frac{\mu}{2T}\right),
\end{equation}
where tunneling current $I$ is given by Eq.~(\ref{Tunneling current at finite temperature. Result.}). This expression is independent of interaction parameter $b$. The current fluctuations satisfy a classical (Poissonian) shot noise form. This is related to uncorrelated tunneling of electrons through a QPC. The result is identical to the case of non-interacting electrons in Landauer-Buttiker approach and shows that the electron-phonon interaction is not important, thus successive electrons tunnel infrequently.  

\section{Tunneling current in ac regime}
\label{Sec:V}
To investigate the time-dependent driven case, we split the voltage applied to the upper edge contact in dc and ac parts, namely
\begin{equation}
V(t)=\mu + \mu_1 \cos(\Omega t),
\end{equation}
where the ac part of $V(t)$ averages to zero over one period $\mathcal{T}=2\pi /\Omega$ and the number of electrons per pulse is $q=\mu/\Omega$. One can consider more exotic shapes of drive, such as "train" of Lorentzian pulses ($q=\mp 1, \mp 2,...$) in context of levitons $-$ the time-resolved minimal excitation states of a Fermi sea, recently detected in 2DEG~\cite{Glattli, Hofer, Rech}.

The time-averaged tunneling current between two edges in the case of ac voltage is given by the formula
\begin{equation}
\label{Definition of current in ac regime}
I=2\text{Re}[\mathcal{I}],
\end{equation} 
where the quantity in square brackets has the form 
\begin{equation}
\label{Complex current in ac regime}
\mathcal{I}=\frac{1}{\mathcal{T}} \int^{\mathcal{T}}_0 dt \int^t_{-\infty} dt^{\prime} e^{i\int^t_{t^{\prime}} V(\tilde{t})d\tilde{t}}\langle [\hat{A}^{\dagger}(t),\hat{A}(t^{\prime})]\rangle,
\end{equation}
and the average is taken with respect to the equilibrium density operator $\rho_0$.
\begin{figure}
\includegraphics[width=8cm]{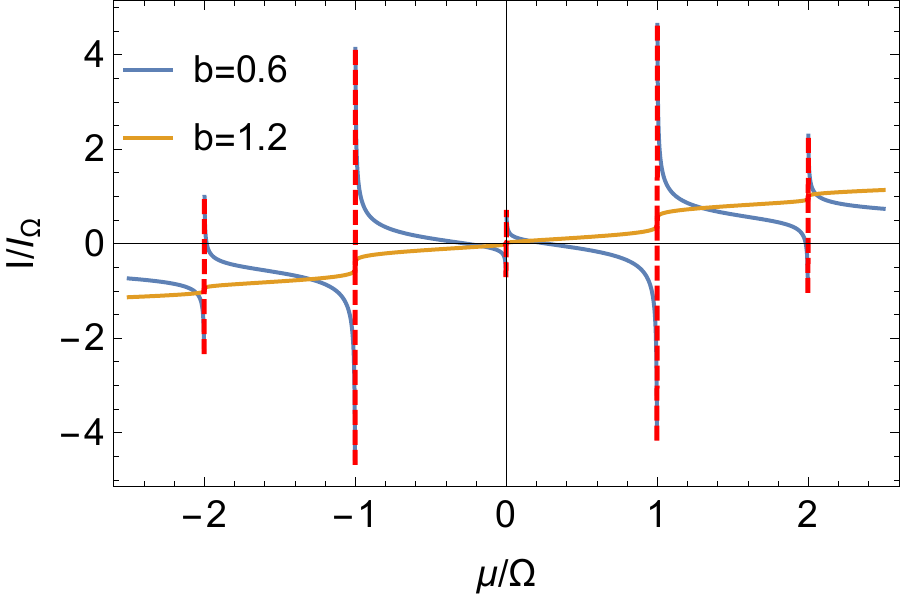}
\caption{\label{fig11.} The normalized tunneling current $I/I_{\Omega}$ is plotted versus the dimensionless dc component of bias $\mu / \Omega$ at zero temperature and fixed $\Omega$. Here we set $b=0.6$, $b=1.2$, $\mu_1 / \Omega =2$ and $I_{\Omega}$ is given by Eq.~(\ref{Tunneling current at zero temperature. Result.}) simply replacing $\mu$ by frequency $\Omega$.}
\end{figure}
After substituting the exact form of vertex operators from Eq.~(\ref{Tunneling Hamiltonian}) into Eq.~(\ref{Complex current in ac regime}) and taking into account that the up and down edges are independent we get  
\begin{widetext}
\begin{equation}
\label{Complex current in ac regime 2}
\mathcal{I}=\frac{1}{\mathcal{T}} \int^{\mathcal{T}}_0 dt \int^t_{-\infty} dt^{\prime} e^{i \mu(t-t^{\prime})}e^{i\frac{\mu_1}{\Omega}[\sin(\Omega t)-\sin(\Omega t^{\prime})]} \left[\langle \hat{\psi}^{\dagger}_u (t) \hat{\psi}_u(t^{\prime})\rangle \langle \hat{\psi}_d (t)\hat{\psi}^{\dagger}_d(t^{\prime})\rangle -  \langle \hat{\psi}^{\dagger}_d (t^{\prime})\hat{\psi}_d(t)\rangle \langle \hat{\psi}_u (t^{\prime})\hat{\psi}^{\dagger}_u(t)\rangle \right].
\end{equation}
\end{widetext}

Further progress in Eq.~(\ref{Complex current in ac regime 2}) is possible using the property of Bessel functions of first kind 
\begin{equation}
\label{Property of Besel function}
\exp[i\xi \sin \varphi]=\sum^{\infty}_{n=-\infty}J_n(\xi)\exp[i n \varphi],
\end{equation}
where $J_n(\xi)$ gives the Bessel function of the first kind and $n$ is an integer.
 
Substituting Eq.~(\ref{Correlation function at zero temperature}) and Eq.~(\ref{Property of Besel function}) into Eq.~(\ref{Complex current in ac regime 2}) and performing the time integration we obtain the final result for tunneling current at zero temperature
\begin{equation}
I \propto \mathcal{D} |\tau|^2 \frac{2\pi}{\Gamma[b]} \sum^{+\infty}_{n=-\infty} J^2_n(\mu_1/\Omega)|\mu+n\Omega|^{b-1}\text{sign}(\mu+n\Omega).
\end{equation}
The normalized tunneling current is plotted in Fig.~\ref{fig11.}. At zero drive frequency, $\mu_1/\Omega \to \infty$ and positive dc bias $\mu$ we recover the result in Eq.~(\ref{Tunneling current at zero temperature. Result.}) of Sec.~\ref{Sec:III}. Because of ac drive here we observe the singularities which strongly depend on the electron-phonon interaction parameter $b$ and are related with a new energy scale, the drive frequency $\Omega$.   
 
Identical steps as in case of zero temperature bring us to the result for finite temperatures, namely
\begin{equation}
\begin{split}
& I \propto 2 \mathcal{D} |\tau|^2 (2\pi T)^{b} \times \\
& \sum^{+\infty}_{n=-\infty} J^2_n(\mu_1 /\Omega)\sinh\left(\frac{\mu+n\Omega}{2T}\right)\frac{\left|\Gamma\left[\frac{1}{2}(b+i\frac{\mu+n\Omega}{\pi T})\right]\right|^2}{2\pi T\Gamma[b]}.
\end{split}
\end{equation}
At zero drive frequency and $\mu_1/\Omega \to \infty$, observing that $\sum^{\infty}_{n=-\infty}J^2_{n}(\mu_1/\Omega) \to 1$, we recover the result provided in Eq.~(\ref{Tunneling current at finite temperature. Result.}) of Sec.~\ref{Sec:III}. We do not provide the plot at finite temperature, which is less interesting compared to the zero temperature case. However, one has to mention that at high temperatures $ T \gtrsim \text{max}\{\mu, \Omega\}$ the singularities are expected to smear out and a thermal broadening is appears. Consequently, the singularities are restored at low temperatures.   

\section{Finite and zero frequency noise in ac regime}
\label{Sec:VI}

\subsection{Finite frequency noise} 
The time-averaged phonon-assisted finite frequency non-symmetrized noise is given by the Wigner transformation defined as
\begin{equation}
\label{Wigner trasnformation finite frequency noise}
S(\omega)=\frac{1}{\mathcal{T}} \int^{\mathcal{T}}_0 d\tau \int d\tau^{\prime} S\left(\tau+\frac{\tau^{\prime}}{2},\tau-\frac{\tau^{\prime}}{2}\right)e^{i\omega \tau^{\prime}},
\end{equation}
where we introduced the "center of mass" $\tau=(t+t^{\prime})/2$ and "relative" $\tau^{\prime}=t-t^{\prime}$ time coordinates. The integrand is given by the current fluctuation correlator in time domain, namely $S(t,t^{\prime})=\langle \delta \hat{J}(t) \delta \hat{J}(t^{\prime})\rangle$ with $\delta \hat{J}(t)=\hat{J}(t)-\langle \hat{J}(t) \rangle$, where the average is taken with respect to ground state of QH edges with free phonons. Eq.~(\ref{Wigner trasnformation finite frequency noise}) is nothing but the Fourier transform of $S(t,t^{\prime})$ in terms of "relative" time coordinate $\tau^{\prime}$.

The straightforward calculations gives the following result for zero temperature case
\begin{equation}
\label{Finite frequency noise at zero temperature in ac regime. Result}
\begin{split}
& S(\omega) \propto \mathcal{D}|\tau|^2 \frac{2\pi}{\Gamma[b]}\times \\
& \sum^{+\infty}_{n=-\infty}\sum_{\sigma=\mp}J^2_n(\mu_1/\Omega)|\omega+\sigma(\mu+n\Omega)|^{b-1}\theta[\omega+\sigma(\mu+n\Omega)].
\end{split}
\end{equation}
\begin{figure}
\includegraphics[width=8cm]{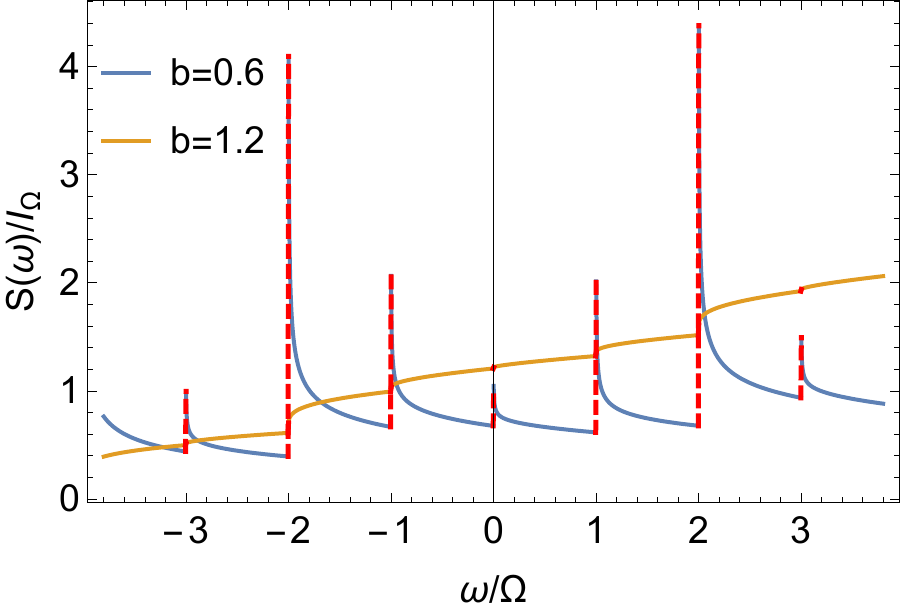}
\caption{\label{fig12.}  The normalized unsymmetrized finite frequency noise $S(\omega)/I_{\Omega}$ is plotted versus the dimensionless frequcny $\omega / \mu$ at zero temperature and fixed $\mu$ and $\Omega$. Here we set $b=0.6$, $b=1.2$, $\mu_1 / \Omega =2$, $\mu / \Omega=3$ and $I_{\Omega}$ is given by Eq.~(\ref{Tunneling current at zero temperature. Result.}) simply replacing $\mu$ by frequency $\Omega$.}
\end{figure}
The normalized finite frequency non-symmetrized noise at zero temperature is plotted in Fig.~\ref{fig12.}.

At finite temperature we get the following result for finite frequency non-symmetrized noise 
\begin{equation}
\label{Finite frequency noise at finite temperature in ac regime. Result}
\begin{split}
& S(\omega) \propto \mathcal{D} |\tau|^2 (2\pi T)^{b}\sum^{+\infty}_{n=-\infty}\sum_{\sigma=\mp} J^2_n(\mu_1 /\Omega) \times \\
& \exp\left[\frac{\omega+\sigma(\mu+n\Omega)}{2T}\right]\frac{\left|\Gamma\left[\frac{1}{2}\left(b+i\frac{\omega+\sigma(\mu+n\Omega)}{\pi T}\right)\right]\right|^2}{2\pi T\Gamma[b]}.
\end{split}
\end{equation} 
At zero frequency we recover the result of next subsection. 
\subsection{Zero frequency noise}
The time-averaged zero frequency non-symmetrized noise is obtained from Eq.~(\ref{Wigner trasnformation finite frequency noise}) and is given by the following anti-commutator of vertex operators
\begin{figure}
\includegraphics[width=8cm]{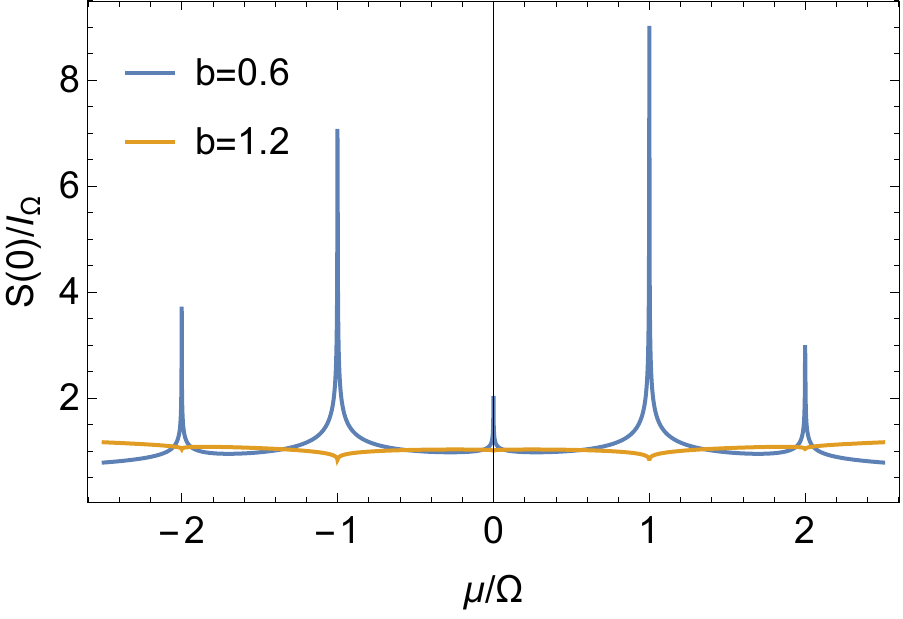}
\caption{\label{fig13.}  The normalized unsymmetrized zero frequency noise $S(0)/I_{\Omega}$ is plotted versus the dimensionless dc component of bias $\mu / \Omega$ at zero temperature and fixed $\Omega$. Here we set $b=0.6$, $b=1.2$, $\mu_1 / \Omega =2$ and $I_{\Omega}$ is given by Eq.~(\ref{Tunneling current at zero temperature. Result.}) simply replacing $\mu$ by frequency $\Omega$.}
\end{figure}
\begin{equation}
\label{Zero frequency noise in ac regime in terms of vertex operators}
S(0)=\frac{1}{\mathcal{T}} \int^{\mathcal{T}}_0 dt \int dt^{\prime} e^{i\int^t_{t^{\prime}} V(\tilde{t})d\tilde{t}}\langle \{ \hat{A}^{\dagger}(t+t^{\prime}/2), \hat{A}(t-t^{\prime}/2)\}\rangle
\end{equation}

At zero temperature and after the time integration in Eq.~(\ref{Zero frequency noise in ac regime in terms of vertex operators}) we obtain the following expression for zero frequency noise
\begin{equation}
\label{Zero frequency noise at zero temperature in ac regime. Result}
S(0) \propto \mathcal{D}|\tau|^2 \frac{2\pi}{\Gamma[b]}\sum^{\infty}_{n=-\infty} J^2_n (\mu_1 /\Omega)|\mu+n\Omega|^{b-1}.
\end{equation}
This result coincides with Eq.~(\ref{Finite frequency noise at zero temperature in ac regime. Result}) for finite frequency noise at zero frequency. The normalized zero frequency noise at zero temperature is plotted in Fig.~\ref{fig13.}.

For finite temperature the straightforward calculation of integral over time variable results in
\begin{equation}
\label{Zero frequency noise at finite temperature in ac regime. Result}
\begin{split}
& S(0) \propto 2 \mathcal{D} |\tau|^2 (2\pi T)^{b} \times \\
& \sum^{+\infty}_{n=-\infty} J^2_n(\mu_1 /\Omega)\cosh\left(\frac{\mu+n\Omega}{2T}\right)\frac{\left|\Gamma\left[\frac{1}{2}(b+i\frac{\mu+n\Omega}{\pi T})\right]\right|^2}{2\pi T\Gamma[b]}.
\end{split}
\end{equation}
Here we can as well mention that the result for finite frequency noise given by Eq.~(\ref{Finite frequency noise at finite temperature in ac regime. Result}) at $\omega \to +0$ transforms into Eq.~(\ref{Zero frequency noise at finite temperature in ac regime. Result}).  

\section{Conclusion}
\label{Sec:VII}
In this paper we have studied the impact of strong electron-acoustic-phonon interaction on
QH edge state transport. We have discussed the case of filling factor $\nu=1$, thus no electron-electron interaction influences on transport properties. Using Wen's effective low-energy theory of QH edge states and applying a BV transformation we were able to take into account the electron-phonon interaction exactly. It was demonstrated that the equilibrium two-point correlation function splits into three independent correlation functions with different velocities. Namely, two of them correspond to "downstream" modes and the third one is related to the "upstream" mode in agreement with Ref.~\cite{Eggert} The presence of electron-phonon interaction strongly modifies the equilibrium propagator, i.e it destroys the Fermi liquid behavior in comparing with non-interacting electrons. It has been already shown that dc Hall conductance of single-branch QH edge with electron-phonon interaction is not varied and is given by quantum conductance, $G_H=e^2/h$~\cite{Eggert}. In current manuscript we state that according to the Wiedemann-Franz law~\cite{Kane, Fisher}, the thermal Hall conductance is not modified as well, namely $K_H=TG_H L_0$, where $L_0$ is a free-fermion Lorentz number.         

We studied the tunneling current for a QPC in lowest order of tunneling coupling under dc and ac biases. Perturbative results are obtained for arbitrary interaction parameter $b$, which depends on electron-phonon coupling, Fermi and sound velocities. We have observed that at high temperatures $\mu/T \ll 1$ the tunneling current is proportional to applied dc bias, $I \propto T^{b-2} \mu$, which is the manifestation of ohmic behavior of QPC. However at low temperatures, $ \mu/T \gg 1$ we get a non-ohmic behavior, $I \propto \mu^{b-1}$, specific for non-Fermi liquid. In contrast, in case of ac drive at zero temperature we observe the singularities in tunneling current. These singularities strongly depend on electron-phonon interaction parameter $b$ and are related with a new energy scale, the drive frequency $\Omega$. One has to mention that at high temperatures $ T \gtrsim \text{max}\{\mu, \Omega\}$ the singularities are expected to smear out and to reveal the thermal broadening. Consequently, at low temperatures the singularities are restored.

Apart from current we have studied as well the non-equilibrium finite frequency non-symmetrized noise of tunneling current in dc and ac regimes. Here the correlations caused by electron-phonon interaction are responsible for algebraic singularities, $\omega=\mp \mu$, $\omega = \mp(\mu+n\Omega)$ at zero temperature, which depend on $b$, the electron-phonon interaction parameter. In case of dc bias the ratio between shot noise and current is independent of $b$ and has the form $S(0)/I=\coth\left(\mu/2T \right)$. Thus, the current fluctuations satisfy a classical (Poissonian) shot noise form. This is related with uncorrelated tunneling of electrons through QPC. The result is identical to the case of non-interaction electrons in Landauer-Buttiker approach and shows that the electron-phonon interaction is not important.  

In conclusion, we have shown that the presence of electron-phonon interaction strongly modifies the analytical structures of noise and current compared to the free-fermion case. As a future perspectives it is appropriate to investigate the tunneling current and noise in two-QPC and resonant level QH systems with Laughlin states in presence of electron-phonon interaction, the periodic train of Lorentzian voltage pulses in context of levitonic physics~\cite{Glattli}.

\acknowledgments
We are grateful to Solofo Groenendijk for careful reading of the manuscript and Thomas L. Schmidt for fruitful discussions. We acknowledge financial support from the Fonds National de la Recherche Luxembourg under Grant No. ATTRACT 7556175 and  INTER 11223315.
 
\appendix
\begin{widetext} 
\section{Eigenvalues of characteristic equation}
\label{Sec:A}
In this section we present the eigenvalues of Eq.~(\ref{Characteristic equation}) from main text. The three distinct real eigenvalues of this cubic equation are given by
\begin{equation}
\begin{split}
& \lambda_{1,2}=\frac{1}{3}\left(\alpha-\frac{1 \pm i\sqrt{3}}{2}\frac{\xi_1}{\sqrt[3]{\xi_2+\sqrt{\xi^2_2-\xi^3_1}}}-\frac{1 \mp i\sqrt{3}}{2}\sqrt[3]{\xi_2+\sqrt{\xi^2_2-\xi^3_1}}\right), \\
& \lambda_3=\frac{1}{3}\left(\alpha+\frac{\xi_1}{\sqrt[3]{\xi_2+\sqrt{\xi^2_2-\xi^3_1}}}+\sqrt[3]{\xi_2+\sqrt{\xi^2_2-\xi^3_1}}\right), 
\end{split}
\end{equation}
where $\xi_1=\alpha^2+3$, $\xi_2=\alpha^3-9\alpha+27\beta^2$ and $\alpha=v_F/v_s$, $\beta=v_c/v_s$. These eigenvalues are important to get the renormalized spectrum of phonons and plasmons $-$ collective excitations in QH edge state (see Fig.~\ref{fig2.} in main text). 

\section{Correlation function at zero temperature}
\label{Sec:B}
In this section we calculate the correlation function of right moving fermions $G(x,t;x^{\prime} t^{\prime})$ at filling factor $\nu=1$ in the presence of electron-phonon interaction at zero temperature. Using the bosonization technique one can write
\begin{equation}
\label{App. Correlation function}
G(x_1 t_1;x_2 t_2) \propto \langle e^{-i \hat{\phi}(x_1,t_1)} e^{i \hat{\phi}(x_2,t_2)} \rangle = e^{M(x_1 t_1;x_2 t_2)}, 
\end{equation}
where in Gaussian approximation and equal bosonic auto-correlators, the function in exponent is given by
\begin{equation}
\label{App. Gaussian approximation}
M(x_1 t_1;x_2 t_2)= \langle \hat{\phi}(x_1,t_1) \hat{\phi}(x_2,t_2) \rangle -  \langle \hat{\phi}^2(x_1,t_1)\rangle.
\end{equation}
Next using the definition of bosonic field in terms of creation and annihilation operators of bosons from Eq.~(\ref{Expansion of bosonic field into creation and annihilation operators}) the cross-correlator of bosonic fields takes the following form (zero modes are omitted)
\begin{equation}
 \langle \hat{\phi}(x_1,t_1) \hat{\phi}(x_2,t_2) \rangle = \sum_{kk^{\prime}>0} \sqrt{\frac{(2\pi)^2}{kk^{\prime} L^2}}\left[ \langle \hat{a}_k(t_1)\hat{a}^{\dagger}_{k^{\prime}}(t_2)\rangle e^{ikx_1-ik^{\prime} x_2}+ \langle \hat{a}^{\dagger}_k(t_1)\hat{a}_{k^{\prime}}(t_2)\rangle e^{-ikx_1+ik_{\prime} x_2}\right].
\end{equation}
Substituting the relation for bosonic operators in terms of BV transformed new bosonic operators from Eq.~(\ref{Bosonic operator after BV transformation}) 
\begin{equation}
\hat{a}_k(t)=b_{11}e^{-iv_1 k t}\hat{\alpha}_k(t)+b_{12}e^{-i v_2 k t}\hat{\beta}_k(t)+b_{13}e^{-iv_3 k t}\hat{\gamma}^{\dagger}_{-k}(t)
\end{equation}
into the above cross-correlator we obtain the following simple combination of sum of three terms  
\begin{equation}
\label{App. Exponent of correlation function}
M(x_1 t_1;x_2 t_2)= b^2_{11}\mathcal{I}_1 (x_1 t_1;x_2 t_2)+b^2_{12}\mathcal{I}_2 (x_1 t_1;x_2 t_2)+b^2_{13}\mathcal{I}_3 (x_1 t_1;x_2 t_2),
\end{equation}
where  
\begin{equation}
\begin{split}
& \mathcal{I}_1 (x_1 t_1;x_2 t_2) = \int^{\infty}_0 \frac{e^{-\eta k}}{k}\left[e^{ik(x_1-x_2)-iv_1 k(t_1-t_2)}-1\right]\left[1+f^{(v_1)}_B(k)\right]+\left[e^{-ik(x_1-x_2)+iv_1 k(t_1-t_2)}-1\right]f^{(v_1)}_{B}(k), \\
& \mathcal{I}_2(x_1 t_1;x_2 t_2) = \int^{\infty}_0 \frac{e^{-\eta k}}{k}\left[e^{ik(x_1-x_2)-iv_2 k(t_1-t_2)}-1\right]\left[1+f^{(v_2)}_B(k)\right]+\left[e^{-ik(x_1-x_2)+iv_2k(t_1-t_2)}-1\right]f^{(v_2)}_{B}(k), \\
& \mathcal{I}_3(x_1 t_1;x_2 t_2) = \int^{\infty}_0 \frac{e^{-\eta k}}{k}\left[e^{ik(x_1-x_2)+iv_3 k(t_1-t_2)}-1\right]f^{(v_3)}_B(k)+\left[e^{-ik(x_1-x_2)-iv_3 k(t_1-t_2)}-1\right]\left[1+f^{(v_3)}_{B}(k)\right].
\end{split}
\end{equation}
Here we introduced the new notations $v_1=\lambda_1 v_s$, $v_2 =\lambda_2 v_s$, $v_3=-\lambda_3 v_s$, where $\lambda_i$, $i=1,2,3$ are eigenvalues of the dynamical matrix and $f^{(v_i)}_{B}=(e^{\beta v_i k}-1)^{-1}$ is a bosonic distribution function, $\beta$ is the inverse temperature and $i=1,2,3$. At zero temperature we have 
\begin{equation}
\label{App. Integrals for three terms}
\begin{split}
& \mathcal{I}_1 (x_1 t_1;x_2 t_2) = \int^{\infty}_0 \frac{e^{-\eta k}}{k}\left[e^{ik(x_1-x_2)-iv_1 k(t_1-t_2)}-1\right]=\log\left[\frac{\eta}{\eta-i[x_1-x_2-v_1 (t_1-t_2)]}\right], \\
& \mathcal{I}_2 (x_1 t_1;x_2 t_2) = \int^{\infty}_0 \frac{e^{-\eta k}}{k}\left[e^{ik(x_1-x_2)-iv_2 k(t_1-t_2)}-1\right]=\log\left[\frac{\eta}{\eta-i[x_1-x_2-v_2 (t_1-t_2)]}\right], \\
& \mathcal{I}_3 (x_1 t_1;x_2 t_2) = \int^{\infty}_0 \frac{e^{-\eta k}}{k}\left[e^{-ik(x_1-x_2)-iv_3 k(t_1-t_2)}-1\right]=\log\left[\frac{\eta}{\eta+i[x_1-x_2+v_3 (t_1-t_2)]}\right].
\end{split}
\end{equation}
Substituting Eq.~(\ref{App. Integrals for three terms}) into Eq.~(\ref{App. Exponent of correlation function}) and then into Eq.~(\ref{App. Correlation function}) we finally get the expression for the two-point correlation function in the form
\begin{equation}
G(x_1,t_1;x_2,t_2) \propto \left[\frac{i\eta/2\pi}{x_1-x_2-v_1(t_1-t_2)+i\eta}\right]^{b^2_{11}} \left[\frac{i\eta/2\pi}{x_1-x_2-v_2(t_1-t_2)+i\eta}\right]^{b^2_{12}} \left[\frac{-i\eta/ 2\pi}{x_1-x_2+v_3(t_1-t_2)-i\eta}\right]^{b^2_{13}},
\end{equation}
where we set $\eta \to \eta/2\pi$ to recover the free fermion case.    
\end{widetext}

\end{document}